\documentclass{ptephy_v1}

\preprintnumber{KEK-TH-2128} 

\usepackage{amsmath,amssymb,bm}

\newcommand{\be}{\begin{equation}}
\newcommand{\ee}{\end{equation}}
\newcommand{\bea}{\begin{eqnarray}}
\newcommand{\eea}{\end{eqnarray}}

\newcommand{\eq}[1]{Eq.~(\ref{eq:#1})}
\newcommand{\sect}[1]{Sec.~\ref{sec:#1}}
\newcommand{\appen}[1]{Appendix~\ref{sec:#1}}

\newcommand{\del}{\partial}

\newcommand{\bra}{\langle}
\newcommand{\ket}{\rangle}

\bmdefine{\bmq}{{\bm{q}}}
\bmdefine{\bmk}{{\bm{k}}}
\bmdefine{\bmx}{{\bm{x}}}
\bmdefine{\bmy}{{\bm{y}}}
\bmdefine{\bmr}{{\bm{r}}}
\bmdefine{\bmnabla}{{\bm{\nabla}}}
\bmdefine{\bmA}{ \bm{A} }
\bmdefine{\bmD}{ \bm{D} }

\bmdefine{\bmPhi}{ \bm{\Phi} }
\bmdefine{\bmPsi}{ \bm{\Psi} }
\bmdefine{\bmcalO}{ \bm{\mathcal{O}} }

\newcommand{\calH}{{\cal H}}
\newcommand{\calI}{{\cal I}}

\newcommand{\calR}{{\cal R}}

\newcommand{\vecx}{\vec{x}}

\newcommand{\nq}{\mathfrak{q}}
\newcommand{\nw}{\mathfrak{w}}

\newcommand{\calP}{{\cal P}}

\bmdefine{\bmg}{{\bm{g}}}
\bmdefine{\bmR}{{\bm{R}}}

\newcommand{\mfh}{\mathfrak{h}}
\newcommand{\mfw}{\nw}
\newcommand{\mfq}{\nq}
\newcommand{\mfQ}{\nq}
\newcommand{\mfA}{\mathfrak{A}}

\newcommand{\dw}{\delta\nw}
\newcommand{\dq}{\delta \nq}
\newcommand{\deta}{\delta\eta}

\newcommand{\twonabla}{\bmnabla}

\newcommand{\odiff}[2]{ \frac{d #1}{d #2} }

\begin{document}

\title{Holographic chaos, pole-skipping, and regularity}


\author[1]{Makoto Natsuume}
\affil[1]{KEK Theory Center, Institute of Particle and Nuclear Studies, High Energy Accelerator Research Organization, Tsukuba, Ibaraki, 305-0801, Japan 
\thanks{Also at 
Department of Particle and Nuclear Physics, 
SOKENDAI (The Graduate University for Advanced Studies), 1-1 Oho, 
Tsukuba, Ibaraki, 305-0801, Japan;
Department of Physics Engineering, Mie University, 
Tsu, 514-8507, Japan.} \email{makoto.natsuume@kek.jp}}

\author[2]{Takashi Okamura}
\affil[2]{Department of Physics, Kwansei Gakuin University, Sanda, Hyogo, 669-1337, Japan \email{tokamura@kwansei.ac.jp}}




\begin{abstract}%
We investigate the ``pole-skipping" phenomenon in holographic chaos. According to the pole-skipping, the energy-density Green's function is not unique at a special point in complex momentum plane. This arises because the bulk field equation has two regular near-horizon solutions at the special point. We study the regularity of two solutions more carefully using curvature invariants. In the upper-half $\omega$-plane, one solution, which is normally interpreted as the outgoing mode, is in general singular at the future horizon and produces a curvature singularity. However, at the special point, both solutions are indeed regular. Moreover,  the incoming mode cannot be uniquely defined at the special point due to these solutions.
\end{abstract}

\subjectindex{AdS/CFT correspondence, Black holes in string theory, Relativity}

\maketitle

\section{Introduction and Summary}

\subsection{Pole-skipping}

In recent years, the quantum many-body chaos attracts much attention. A useful probe of chaos is out-of-time-ordered correlation function (OTOC). An OTOC shows the early-time exponential growth for a chaotic system:
\be
C(t,\vecx) = \bra V(t,\vecx) W(0) V(t,\vecx) W(0) \ket_\beta \simeq 1 - e^{\lambda(t-x/v_B)} + \cdots~,
%
\ee
where $V$ and $W$ are generic operators, $\beta$ is the inverse temperature. $\lambda$ is the (quantum) Lyapunov exponent and $v_B$ is the butterfly velocity. Generically, $\lambda$ satisfies the bound \cite{Maldacena:2015waa}
\be
\lambda \leq 2\pi T~.
\label{eq:chaos_bound}
\ee

The AdS/CFT duality or holography \cite{Maldacena:1997re,Witten:1998qj,Witten:1998zw,Gubser:1998bc} is a useful tool to study quantum many-body systems (see, {\it e.g.}, Refs.~\cite{CasalderreySolana:2011us,Natsuume:2014sfa,Ammon:2015wua,Zaanen:2015oix,Hartnoll:2016apf}). It is conjectured that a holographic system saturates the bound, or black hole is maximally chaotic \cite{Shenker:2013pqa,Roberts:2014isa,Shenker:2014cwa,Maldacena:2015waa}. 

The quantum chaos has been studied using OTOC, but recently it is claimed that the chaotic behavior can be seen even at the level of retarded Green's functions. This phenomenon is known as ``pole-skipping" \cite{Grozdanov:2017ajz,Blake:2018leo}%
\footnote{See, {\it e.g.}, Refs.~\cite{Gu:2016oyy,Haehl:2018izb} for discussion of the pole-skipping from field theory point of view.}.
To motivate the pole-skipping, write $C(t,\vecx)$ in a plane-wave form
\be
C(t,\vecx) \simeq 1-e^{-i\omega t+iqx}
%
\ee
with purely imaginary values of $(\omega, q)$:
\be
\omega = i\lambda =: \omega_\star~, \quad q=i\frac{\lambda}{v_B} =: q_\star~.
\label{eq:special_pt}
\ee
For Schwarzschild-AdS$_{p+2}$ black holes \cite{Shenker:2013pqa}, 
\be
\lambda = 2\pi T~, \quad v_B^2=\frac{p+1}{2p}~.
%
\ee

The pole-skipping claims that retarded Green's function shows a characteristic behavior at the ``special point" in momentum space $(\omega_\star,q_\star)$. More explicitly, consider the energy-density 2-point function. Any perturbation carries energy, so one would expect to see the chaotic behavior from the energy-density 2-point function. Generically, one would write the function as
\begin{align}
G^R_{T^{00}T^{00}} (\omega,q) = \frac{b(\omega,q)}{a(\omega,q)}~.
\label{eq:2pt}
\end{align}
The pole-skipping claims that 
\begin{align}
a(\omega_\star,q_\star) = b(\omega_\star,q_\star) = 0~,
%
\end{align}
and one can locate $\lambda$ and $v_B$ in this way. Then, naively $G^R=0/0$, but more precisely, $G^R$ is not uniquely determined at the special point. Near the special point, 
\begin{align}
G^R = \frac{ \delta\omega (\del_\omega b)_\star + \delta q (\del_q b)_\star +\cdots }{\delta\omega (\del_\omega a)_\star + \delta q (\del_q a)_\star +\cdots }
= \frac{ (\del_\omega b)_\star + \frac{\delta q}{\delta\omega} (\del_q b)_\star +\cdots }{ (\del_\omega a)_\star + \frac{\delta q}{\delta\omega} (\del_q a)_\star +\cdots }~.
%
\end{align}
Then, the Green's function at the special point is not unique because it depends on the slope $\delta q/\delta\omega$.

The pole-skipping is interesting and is useful. Perviously, one needs to evaluate an OTOC or a 4-point function in order to see a chaotic behavior. But a real-time finite-temperature 4-point function is not easy to evaluate even in the AdS/CFT duality. In principle, one can compute them \cite{Skenderis:2008dh}, but such a computation is extremely rare. As a result, OTOC has not been extensively studied in a variety of bulk systems. One approach is to use the WKB approximation which assumes large scaling dimensions for $V$ and $W$ \cite{Shenker:2013pqa}. As another approach, one can compute OTOC in the AdS$_2$ black hole \cite{Jensen:2016pah} since one can map the black hole to the pure AdS$_2$ via a conformal transformation. But the chaos bound \eqref{eq:chaos_bound} has not been explicitly verified in a variety of bulk systems. 
This situation contrasts with the viscosity bound $\eta/s=1/(4\pi)$ \cite{Kovtun:2004de}, 
which has been extensively verified in a variety of bulk systems. 

However, the pole-skipping claims that the chaotic behavior can be seen at the level of 2-point functions, and these are objects we often compute in the AdS/CFT duality. So, the approach allows us to study the holographic chaos more in details in a variety of bulk systems. However, the pole-skipping has many unanswered questions. Most importantly, it is not clear why the 2-point function has anything to do with the 4-point function.

\subsection{Universality and near-horizon physics}

It is conjectured that holographic systems saturate the chaos bound. From the bulk point of view, such a universal behavior is often related to the universal nature of the near-horizon physics. One well-known example is $\eta/s$. So, one would expect that the near-horizon physics plays an important role in the holographic chaos as well. Ref.~\cite{Blake:2018leo} studies this issue in the context of the pole-skipping. 

The energy-density 2-point function corresponds to solving the scalar mode (sound mode) of gravitational perturbations, so they examine the perturbation problem. They found that the near-horizon physics shows a special behavior at the special point, namely the special point is characterized by the appearance of an extra regular incoming solution.
The field equation in general has 2 solutions, the incoming mode and the outgoing mode. We are interested in the retarded Green's function, so we choose the incoming mode. However, at the special point, 
\begin{itemize}
\item Both solutions are regular at the horizon. 
\item One cannot distinguish between the incoming and the outgoing modes. 
\end{itemize}
Then, the incoming mode is not uniquely defined, and as a result, the Green's function is not unique. 

They conclude the regularity of the extra solution just by looking at the mode function, but this definition is ambiguous. Even if a mode function diverges at the horizon, multiplying an appropriate function which vanishes at the horizon can produce a new mode function which is regular. Since the regularity is a key feature of the pole-skipping, it is worthwhile to study the regularity more in details. This is a purpose of this paper. 

The regularity of an arbitrary mode function is not meaningful. From the bulk point of view, we should require a mode function not to produce a curvature singularity. We use ``regularity" in this sense. We reanalyze the scalar mode in the SAdS$_4$ black hole background using variables where regularity can be clearly seen. We also compute curvature invariants for perturbations. 

In a perturbation problem, it is often useful to use a ``master variable." But one needs to be careful since it may fail at particular points in momentum space. Instead, it is straightforward to analyze the special point if one uses the full set of gauge-invariant variables and the full linearized Einstein equation. This allows one to analyze the other systems where master variables are hard to find. 

Our results are summarized as follows:%
\footnote{While this paper and the companion paper \cite{Natsuume:2019xcy} are in preparation, there appeared preprints \cite{Grozdanov:2019uhi,Blake:2019otz} which have some overlap with ours. } 
\begin{itemize}
\item
At the special point, two solutions are indeed regular, and the curvature remains finite. 
\item
The outgoing mode with $\Im\omega>0$ is in general singular and produces a curvature singularity at the future horizon. This excludes the existence of any other special points in the upper-half $\omega$-plane. (In this paper, we focus on the upper-half $\omega$-plane.)
\item
The field equation has the regular singularity at the horizon $r=1$, but at the special point, it becomes a regular point in the incoming Eddington-Finkelstein (EF) coordinates. As a result, two regular solutions appear.
\end{itemize}
As we see below, the incoming-wave boundary condition is not uniquely defined at the special point. The boundary condition is imposed by deviating from the special point and taking the $\delta\omega, \delta q\to 0$ limit. But the limit is not unique and has a slope dependence $\delta\omega/\delta q$. The existence of two regular solutions reflects the slope dependence.

\section{Sound mode}

We consider the 4-dimensional pure gravity%
\footnote{We use upper-case Latin indices $M, N, \ldots$ for the 4-dimensional bulk spacetime coordinates and use Greek indices $\mu, \nu, \ldots$ for the 3-dimensional boundary coordinates. The boundary coordinates are written as  $x^\mu = (t, x^i)=(t,x,y,\cdots)$. Lower-case Latin indices $a,b,\cdots$ are used for the 2-dimensional subspace $(v,r)$.}
\be
S = \frac{1}{16\pi G_4} \int d^4x \sqrt{-g} (R-2\Lambda)~, \quad \Lambda=-\frac{3}{L^2}~,
%
\ee
and consider the Schwarzschild-AdS$_4$ (SAdS$_4$) black hole:
\begin{align}
ds^2 &= r^2(-fdt^2+d\vecx_2^2) + \frac{dr^2}{r^2f}~,\\
f &= 1-r^{-3}~.
%
\end{align}
For simplicity, we set the AdS radius $L=1$ and the horizon radius $r_0=1$. The Hawking temperature is given by $2\pi T=3/2$.

We solve perturbations in the black hole background. As usual, we impose the incoming-wave boundary condition at the horizon. Using the tortoise coordinate $dr_*:=dr/(r^2f)$, an incoming wave behaves like $e^{-i\omega(t+r_*)}$, and an outgoing wave behaves like $e^{-i\omega(t-r_*)}$. Thus, it is convenient to set $v=t+r_*$ and work with the incoming Eddington-Finkelstein (EF) coordinates. The metric becomes
\be
ds^2 = r^2(-fdv^2+d\vecx_2^2) +2dvdr~.
%
\ee
In the EF coordinates, an incoming wave behaves like $e^{-i\omega v}$, and an outgoing wave behaves like
\begin{align}
e^{-i\omega v} e^{2i\omega r_*} \simeq e^{-i\omega v} (r-1)^{2i\omega/f'(1)} = e^{-i\omega v} (r-1)^{i\omega/(2\pi T)}~.
%
\end{align}
The horizon consists of the future horizon and the past horizon. We use the incoming EF coordinates, so the horizon $r=1$ corresponds to the future horizon (for a finite $v$).

\subsection{Remarks on incoming/outgoing modes}

In the pole-skipping, the issue of the incoming and outgoing modes is rather confusing, so it would be helpful to make a few remarks. 

\paragraph{Definition of incoming/outgoing mode}
At the special point, $\omega,q$ are pure imaginary, and the incoming (outgoing) nature is somewhat obscure, so let us first start from the definition of these modes. 

The incoming and outgoing modes at the future horizon behave like $e^{-i\omega(t\pm r_*)}$. This means that we first define them for $\omega\in\mathbb{R}$. In a black hole background, we are also interested in $\omega\in\mathbb{C}$, and the incoming and outgoing modes are defined by analytic continuation from $\omega\in\mathbb{R}$. This definition seems to work fine in most cases. Then, one would conclude that the distinction between 2 modes is clear even at the special point. However, as we see below, one cannot distinguish between the incoming and the outgoing modes at the special point. 

\paragraph{Importance of boundary condition}
To clarify our statement about the pole-skipping, let us make a few {\it incorrect} statements. If one said that 
\begin{center}
``there is an extra regular solution at the special point, so one must include it," 
\end{center}
this is not really correct. What solution one chooses depends on the boundary condition. If one just considers the general solution, one does not impose a boundary condition. Another misleading statement is 
\begin{center}
``the outgoing mode is also regular at the special point, so one must include it." 
\end{center}
Again, whether one selects the outgoing mode or not depends on the boundary condition. According to the standard AdS/CFT rule, the choice of boundary condition at the horizon reflects the choice of Green's function. The incoming mode corresponds to the retarded Green's function. If one is interested in the retarded Green's function, one should not include the outgoing mode. Regularity is just a prerequisite. 

Therefore, what matters eventually is the incoming-wave boundary condition at the special point. As we discuss below, the incoming-wave boundary condition is not uniquely defined [see \eq{BC_sound}], or the incoming mode is not uniquely determined at the special point. This is the reason why we have to include both solutions. One solution is normally interpreted as an outgoing mode. We exclude it since we normally compute the retarded Green's function. It is excluded from the point of view of regularity as well. But at the special point, it is also regular. Moreover, it should be taken into account to define the incoming mode. 

\paragraph{Advanced Green's function?}
The outgoing mode is in general prohibited from regularity, but it is prohibited at the future horizon $\calH^+$. 
Again, according to the standard AdS/CFT rule, the choice of a boundary condition at the horizon simply reflects the choice of Green's function. The incoming (outgoing) mode at the future horizon corresponds to the retarded (advanced) Green's function. So, one would argue that these modes must be treated symmetrically and the outgoing mode should also be allowed. 

The asymmetry comes from the incoming EF coordinates. We use the incoming EF coordinates, so we impose the boundary condition at the future horizon $\calH^+$. In order to compute the advanced Green's function, one should impose a boundary condition at the past horizon $\calH^-$, not at the future horizon $\calH^+$. In other words, one should use the outgoing EF coordinates. The outgoing mode is allowed at $\calH^-$. 



\subsection{Gauge-invariant variables}

We consider gravitational perturbations of the form
\begin{align}
h_{MN}(r)\, e^{-i\omega v +iqx}~.
%
\end{align}
As is well-known, gravitational perturbations are decomposed as scalar mode, vector mode, and tensor mode. (For $p=2$, there is no tensor mode though.)  
The perturbations are decomposed under the transformation of boundary spatial coordinate $x^i$ (see \appen{gauge_inv_variables} for the details). For example, the scalar mode transforms as scalar under the transformation. 
In this paper, we are interested in the energy-density 2-point function. This corresponds to solving the scalar mode, which has 7 perturbations. For $p=2$, they are given by
\begin{align}
h_{vv}~, h_{vr}~, h_{rr}~, h_{vx}~, h_{rx}~, h_{xx}~, h_{yy}~.
%
\end{align}

The scalar mode has 7 perturbations, but they are redundant due to the diffeomorphism. Normally, one fixes the gauge $h_{rM}=0$, which reduces to 4 perturbations. Then, one constructs gauge-invariant variables which are invariant under the residual gauge transformation. This is the formalism advocated {\it e.g.}, by Kovtun and Starients \cite{Kovtun:2005ev}. 

Instead, we do not fix the gauge and carry out analysis in a fully gauge-invariant manner. We essentially follow the formalism by Kodama and Ishibashi \cite{Kodama:2003jz} and use the variables which are invariant under the full diffeomorphism%
\footnote{In the terminology of Kodama and Ishibashi, the scalar mode here is called ``scalar-type" perturbations.}. 
In either case, there are 4 variables. We denote the gauge-invariant variables as $\mfh_{vv}, \mfh_{vr}, \mfh_{rr}$, and $\mfh_L$ defined by Eqs.~\eqref{eq:scalar-mfh-all}.

So far, we use only the gauge invariance to reduce degrees of freedom. We now impose the equations of motion. Then, these 4 gauge-invariant variables are not independent, and the equations of motion leave us only a single degree of freedom which obeys a second-order differential equation. They are referred as the master field and the master equation. The linearized Einstein equation reduces to
\begin{subequations}
\label{eq:EOM-mfh}
\begin{align}
   0
  &= \mfh_{rr} + \frac{2}{r^2\, f}\, \mfh_{vr}
  ~,
\label{eq:EOM_const0} \\
   0
  &= 2\, \left\{ \frac{ (i\, \mfw - r)\,
      \left( i\, \mfw\, r - \mfQ^2 \right) }{r^2\, f}
      + i\, \mfw - \frac{\mfQ^2}{3\, r}
    \right\}\, \mfh_{vv}
  + \left( \mfw^2 - \mfQ^2 + \mfQ^2\, \frac{f}{3} \right)\,
    r^3\, f\, \mfh_{rr}
\nonumber \\
  &
  - i\, \mfw\, \left\{ \frac{3}{f}\, \left( 1 + \frac{\mfw^2}{r^2} \right)
    - 2\, \left( 2 + \frac{\mfQ^2}{r^2} \right) + f \right\}\, \mfh_L
  ~,
\label{eq:EOM_const1} \\
   0
  &= \mfh'_{vv}
  - \left( 1 - \frac{\mfQ^2}{i\, \mfw}\, \frac{1}{r } \right)\,
    \frac{ \mfh_{vv} }{r}
  - \frac{3}{2}\, \left( \frac{i\, \mfw}{2}
      + \frac{ \mfQ^2 }{i\, \mfw}\, \frac{f}{3}
    \right)\, r^2\, f\, \mfh_{rr}
  - \frac{3}{2\, r^4}\, \mfh_L
  ~,
\label{eq:EOM-hvv} \\
   0
  &= \mfh'_{rr}
  + 3\left\{ 2 - \frac{i\mfw}{r} 
  + \left( \frac{\mfq^2+i\mfw r}{r(r-i\mfw)} - 2 \right)\frac{f}{3} - \frac{\mfq^2}{i\mfw(r-i\mfw) } \left( \frac{f}{3} \right)^2 \right\}
\frac{\mfh_{rr}}{rf} 
\nonumber \\
&\hspace*{0.5truecm}
- \frac{1+2\mfq^2r+3i\mfw r^2}{r-i\mfw} \frac{\mfh_{L}}{r^7f}  + \frac{ 2(\mfq^2-3i\mfw r) }{ 3i\mfw(r-i\mfw) } \frac{\mfh_{vv}}{r^5f}~,
\label{eq:EOM-hrr} \\
   0
  &= \mfh'_{L}
  - \left( 1 + \frac{i\, \mfw}{r} + \frac{f}{3} \right)\,
    \frac{3}{2\, f}\,  \frac{ \mfh_{L} }{r}
  - \left( 1 - \frac{\mfQ^2}{i\, \mfw}\, \frac{1}{r} \right)\,
    \frac{1}{f}\, \frac{ \mfh_{vv} }{r}
  - \frac{\mfQ^2}{2\, i\, \mfw}\, r^2\, f\, \mfh_{rr}
  ~,
\label{eq:EOM-hL}
\end{align}
\end{subequations}
where $'=\del_r$ and
\begin{align}
\nw= \frac{\omega}{2\pi T}~, \quad \nq = \frac{q}{2\pi T}v_B = \frac{\sqrt{3}}{2}\frac{q}{2\pi T}~.
%
\end{align}
In terms of $\nw$ and $\nq$, the special point is located at $(\nw_\star, \nq_\star)=(i, \pm i)$.

There are 2 constraint equations which do not involve $r$-derivatives and 3 differential equations which have one $r$-derivative. The latter 3 are not independent; one is redundant from the constraint equation \eqref{eq:EOM_const1}. The constraint equations \eqref{eq:EOM_const0} and \eqref{eq:EOM_const1} allow us to choose 2 independent variables. Both obey first-order differential equations, so one gets a second-order differential equation for a single variable which is the master equation.

Thus, there is only a single degree of freedom, but note the choice of a master field is not unique. One can choose any of 4 gauge-invariant variables $\mfh_{vv}, \mfh_{vr}, \mfh_{rr}$, and $\mfh_L$ and linear combinations as a master field. 
Among them, a few choices are worth mentioning:
\begin{itemize}
\item From the boundary point of view, it is natural to choose a master variable which does not involve $r$-derivatives of metric perturbations since one imposes the Dirichlet boundary condition at infinity. This is the choice, {\it e.g.}, by Kovtun and Starinets \cite{Kovtun:2005ev}%
\footnote{Note that our gauge-invariant variables implicitly depend on $h'_{MN}$ through $\eta_a$ [see \eq{scalar-mfh-all}], so a general linear combination depends on $h'_{MN}$.}. 
\item It is often useful to rewrite the master equation in the form of a Schr\"{o}dinger equation. This is the choice of Ref.~\cite{Blake:2018leo}.
\end{itemize}

\subsection{A criterion of regularity}

We are interested in the regularity of the perturbations, and we eventually show this from geometric quantities. Any variable is fine in principle. But it would be better if one could check regularity from the behavior of the mode function. So, it is worthwhile to pause here and to consider which variable is suitable for that purpose. 

As mentioned in Introduction, an arbitrary mode function is not very suitable. A diverging mode function can be regular by multiplying an appropriate power of $(r-1)$. One useful criterion is as follows:
%
%
\begin{center}
In the incoming EF coordinates, all gauge-invariant metric perturbations \\
must be smooth (more precisely $C^\infty$) at the future horizon $\calH^+$.
\end{center}
Unlike the Schwarzschild coordinates, the incoming EF coordinate system is regular at the future horizon. 
Thus, the metric perturbations must be smooth there as well. Then, the Riemann tensor components of the perturbed spacetime are also smooth. The master variable of Ref.~\cite{Blake:2018leo} is not appropriate for that purpose. It is not a metric perturbation itself, and it is unclear if all metric perturbations are regular or not.

As we see below, gauge-invariant metric perturbations are expanded as a Taylor series for the $\calH^+$-incoming mode. On the other hand, perturbations are not a Taylor series and curvature invariants diverge for the $\calH^+$-outgoing mode. 

\subsection{Near-horizon analysis (generic)}

Before we solve the sound mode at the special point, let us solve the problem for a generic $(\nw, \nq)$ as a warmup exercise. The choice of the master variable is not unique, but we are interested in regularity of all metric perturbations. So, it is better to choose a variable whose regularity guarantees the regularity of the other variables. The variable $\mfh_{rr}$ is most suitable for the purpose. The other variables are expressed by $\mfh_{rr}$ and $\mfh'_{rr}$ (\appen{relation_mfh}), and the $\mfh_{rr}$ regularity at the horizon guarantees the regularity of $\mfh_{vv}$, $\mfh_{vr}$, and $\mfh_{L}$.
However, this fails when $i\nw+\nq^4=0$ which includes the special point $(\nw,\nq)=(i,\pm i)$, so the special point must be examined separately. 

The $\mfh_{rr}$-equation becomes
\begin{align}
   0
  &= \mfh''_{rr}
  + \left( \frac{2}{r} + \frac{f'}{f}
    - 3\, \frac{ i\, \mfw - 2\, r }{r^2\, f} \right)\, \mfh'_{rr}
  + \frac{3}{r^2\, f}\, \left( 4 - 3\, \frac{i\, \mfw}{r}
    - \frac{\mfQ^2}{r^2} \right)\, \mfh_{rr}
  ~.
\label{eq:EOM-h_rr-only}
\end{align}
The equation has regular singularities at $r=0, \infty,$ and at 3 zeros of $f$ which includes $r=1$. According to Ref.~\cite{Blake:2018leo}, the near-horizon behavior is important for the pole-skipping, so solve the equation by a power series expansion around $r=1$:
\begin{align}
\mfh_{rr} = (r-1)^\lambda \sum_{n=0}\, a_n\, (r-1)^n~.
%
\end{align}
At the lowest order, one gets the indicial equation and obtains
\begin{align}
\lambda_1=0~, \quad \lambda_2 = -2+i\nw~.
%
\end{align}
The coefficient $a_n$ is obtained by a recursion relation. The $\lambda_1$-mode is the incoming mode since $\mfh_{rr} \propto e^{-i\omega v}$, and the $\lambda_2$-mode is the outgoing mode. We show regularity  carefully from curvature invariants later (\sect{curv_inv}). But as discussed in the previous subsection, one may check regularity from $\mfh_{rr}$. The $\lambda_1$-mode is regular at the horizon, but the $\lambda_2$-mode is singular at the horizon in general when $\Im\nw>0$. This result does not apply to the special point though since the master variable $\mfh_{rr}$ fails there. 

One needs a slight modification when $\lambda_1$ and $\lambda_2$ differ by an integer. 
In such a case, the smaller root fails to produce the independent solution since the recursion relation breaks down at some $a_n$. Suppose that the smaller value is $\lambda_2$. Write 2 solutions as $\mfh_1$ (for $\lambda_1$) and $\mfh_2$ (for $\lambda_2$). Then, the second solution in general takes the form
\begin{align}
\mfh_2 = \mfh_1 \ln(r-1) + (r-1)^{\lambda_2} \sum_{n=0}\, b_n\, (r-1)^n~.
%
\end{align}
In any case, the leading behavior near the horizon comes from the second term when $\Im\nw>0$, and the $\lambda_2$-mode is again singular in general.

\section{Special point}
 
\subsection{Solution at special point}

The master variable is a useful technique, but it has some problems:
\begin{itemize}
\item First, it is often not easy to find a master variable. 
\item Second, the choice of the master variable is not unique, so one needs to find which variable is most suitable. This poses a problem particularly for the pole-skipping since some master variable like $\mfh_{rr}$ breaks down at the special point so should not be used there. 
\end{itemize}
In order to analyze the special point, it is straightforward to use the full linearized Einstein equation \eqref{eq:EOM-mfh} instead of a master equation. At the special point, the linearized Einstein equation becomes
\begin{subequations}
\label{eq:EOM-mfh-sp}
\begin{align}
  & 0
  = 2\, r^2\, \mfh_{vv}
  - (r^2 + r + 1)^3\, \mfh_{rr}
  - 3\, (r + 1)\, \mfh_L
  ~,
\label{eq:EOM-const1-sp} \\
  & 0
  = \mfh_{vv}' - \frac{1}{r + 1}\, \mfh_{vv}
  + \frac{ (r + 2)\, (r^2 + r + 1)\, (r^3 + 3\, r + 2) }
         { 4\, r^3\, (r + 1) }\, \mfh_{rr}
  ~,
\label{eq:EOM-hvv-sp} \\
  & 0
  = \mfh'_{rr}
  + 3\, \frac{ 1 + r }{ 1 + r + r^2 }\, \mfh_{rr}
  ~,
\label{eq:EOM-hrr-sp} \\
  & 0
  = \mfh'_{L} - \frac{2}{r}\, \mfh_{L}
  - \frac{ (r + 2)\, (r^2 + r + 1) }{2}\,
    \mfh_{rr}
  ~.
\label{eq:EOM-hL-sp}
\end{align}
\end{subequations}
For a generic $(\nw,\nq)$, the linearized Einstein equation has a regular singularity at the horizon at $r=1$. However, at the special point, the regular singularity becomes a regular point. 
Namely,
\begin{center}
The special point is characterized by the regular singularity at the horizon $r=1$ \\
becoming a regular point in the incoming EF coordinates.
\end{center}
As a result, two independent solutions both become regular. 

One can obtain the solution explicitly at the special point%
\footnote{Actually, one can obtain the general solution not only for the special point but also for $i\nw+\nq^4=0$.\label{fnote:solution}}. 
In this particular example, $\mfh_{rr}$ is independent from the others and obeys a first-order differential equation, so one can solve it as
\begin{subequations}
\label{eq:sol-sound}
\begin{align}
  \mfh^\star_{rr}
  &= C_1\, \calR(r)
  ~,
\label{eq:sol-hrr-sp} \\
  \calR(r)
  &:= \left( \frac{3}{ r^2 + r + 1 } \right)^{3/2}\,
  \exp\left\{ - \sqrt{3}~\left(
    \text{Arctan} \frac{ 2\, r + 1 }{\sqrt{3} } - \frac{\pi}{3} \right)
  \right\} \\
  &\sim C_1\{ 1-2(r-1)+\cdots \} \quad (r\to1)
  ~.
\label{eq:def-calR-special_pt} 
\end{align}
%
Then, the rest is solved as
%
\begin{align}
  & \mfh^\star_{L}
  = r^2\, \left\{\, C_1\, \calI(r) + C_2 \right\}
  ~,
\label{eq:sol-hL-special} \\
  & \mfh^\star_{vv}
  = \frac{3}{2}\, (r + 1)\, \Bigg[\, C_1\,
      \left\{ \frac{ (r^2 + r + 1)^3 }{3\, r^2\, (r + 1)}\,
        \calR
    + \calI(r) \right\}
  + C_2\, \Bigg]
  ~,
\label{eq:sol-hvv-special} \\
  & \calI(r)
  := \int^{r}_1 dr'~
    \frac{ (r' + 2)\, (r'^2 + r' + 1) }{2\, r'^2}\,
    \calR(r')
  ~.
\label{eq:def-intI}
\end{align}
\end{subequations}
Two solutions are indeed regular at the horizon. The $\mfh_{rr}$-solution depends only on 1 integration constant $C_1$. This is rather unusual, but the full solution still depends on 2 integration constants $C_1, C_2$. We consider linear perturbations, so an overall constant is not relevant, but the solution is not unique  due to $C_2/C_1$. 

The above solution is the general solution, and we have not imposed a boundary condition at the horizon. The field equation has two independent solutions. Normally, one is the incoming mode, and the other is the outgoing mode. We usually pick up the incoming mode to compute the retarded Green's function. 

At the special point, the regular singularity becomes a regular point. As a result, two solutions are both regular. Does this mean that there exists a regular outgoing mode? Below we argue that the incoming mode is not uniquely determined at the special point. 

\subsection{Expansion around special point}

For the right interpretation, move away from the special point:
\begin{subequations}
\label{eq:exp-mfw-special-all}
\begin{align}
  & \mfw = \mfw_\star + \delta\mfw = i + \delta\mfw
  ~,
  & \mfQ = \mfQ_\star + \delta \mfQ = \pm\, i + \delta \mfQ
  ~,
\label{eq:exp-mfw-special} \\
  & \mfh_{MN} = \mfh^\star_{MN} + \delta\mfh_{MN} 
  ~.
\label{eq:exp-mfh-special_pt}
\end{align}
\end{subequations}
Instead of $\nq$, it is convenient to use the combination 
\begin{align}
  & \eta := \frac{\mfQ^2}{i\, \mfw}~.
\label{eq:def-eta-delta_eta}
\end{align}

Away from the special point, the field equations have a regular singularity at $r=1$ as usual, and the distinction between the incoming mode and the outgoing mode should be clear. Thus, we move away from the special point and approach the special point $\dw,\dq\to0$. In this way, one expects to obtain the incoming mode at the special point. However, as we see below, one cannot uniquely determine the incoming mode at the special point since the incoming mode depends on the slope $\dw/\dq$ how one approaches the special point.

The resulting equations are rather lengthy, so we present them in \appen{perturbed_eqs_app}, but for example, the $\delta\mfh'_{L}$-equation is given by
\begin{subequations}
\begin{align}
   0
  &= \delta\mfh'_{L}
  - \frac{1}{2}\, \left( 1 + \frac{3\, r^2}{r^2 + r + 1} \right)
    \, \frac{ \delta\mfh_{L} }{r}
  - \frac{r}{r^2 + r + 1}\, \delta\mfh_{vv}
  - \frac{1}{2}\, r^2\, f\, \delta\mfh_{rr}
\nonumber \\
  &\hspace*{0.5truecm}
  + \frac{3\, i\, \delta\mfw}{2\, r^2\, f}\, \Bigg[\,
      \frac{\delta\eta}{i\, \delta\mfw}\,
      \frac{2\, r}{r^2 + r + 1}\, \mfh^\star_{vv}
    + \left\{ -1 + \frac{\delta\eta}{i\, \delta\mfw}\,
      \frac{ r^4\, (r + 1)\, f^2 }{ (r^2 + r + 1)^3 } \right\}
      \, \mfh^\star_{L}
  \, \Bigg]
  ~.
\tag{\ref{eq:exp-EOM-hL-sp}}
\end{align}
\end{subequations}
It may look complicated, but the structure is simple: the perturbation obeys an inhomogeneous differential equation, and the source is given by the special point solution $\mfh^\star_{MN}$. We are interested in the retarded Green's function, so we impose the incoming-wave boundary condition on the perturbations $\delta\mfh_{MN}$.  An incoming wave is written as a Taylor series in the incoming EF coordinates, so the homogeneous part must be expanded as a Taylor series. However, the source term is proportional to $(r-1)^{-1}$, so the equation in general produces an outgoing mode. To avoid this, we require that the source term is also written as a Taylor series. 

From \eq{exp-EOM-hL-sp} and \eq{EOM-const1-sp}, we obtain conditions for the special point solution:
\begin{align}
  & \left. \frac{\mfh^\star_L}{ \mfh^\star_{vv} }\, \right|_{\calH^+}
  = \frac{2}{3}\, \frac{\delta\eta}{i\, \delta\mfw}
  ~,
& & \left. \frac{ \mfh^\star_{rr} }{ \mfh^\star_{vv} }
  \, \right|_{\calH^+}
  = \frac{2}{27}\, \left( 1 - \frac{2\, \delta\eta}{i\, \delta\mfw} \right)
  ~.
\label{eq:BC_sound}
\end{align}
This is the incoming-wave boundary condition for $\mfh^\star_{MN}$. The boundary condition does not uniquely determine $\mfh^\star_{MN}$ and depends on the slope $\deta/\dw$. The exact solution we obtained depends on the combination $C_2/C_1$. This reflects the slope dependence. Conversely, given a $\deta/\dw$, one has to choose the combination $C_2/C_1$ appropriately. 
Imposing the boundary condition on the special point solution \eqref{eq:sol-sound}, we obtain
%
\begin{align}
  & \frac{C_2}{C_1}
  = \frac{-9\, \gamma}{2\, \gamma - 1}
  =: c^\star~,
\label{eq:scalar_zeroth_sol-cond-by_C}
\end{align}
where $\gamma:=\deta/(i\dw)$. 

We emphasized the gauge-invariant variables $\mfh^\star_{MN}$ in this paper, but in order to obtain the Green's function, it is convenient to use a master variable \`{a} la Kovtun and Starinets \cite{Kovtun:2005ev}. One imposes the Dirichlet boundary condition asymptotically, and the master variable does not involve $r$-derivatives of metric perturbations. The master variable $Z_G$ is written in terms of our variables $\mfh_{MN}$ as 
\begin{align}
   Z_G
  &= \frac{1}{ \bmg_{xx} }\, \left( \mfh_{vv}
    - \frac{ \bmg_{vv}' }{ \bmg_{xx}' }\, \mfh_L \right)~.
\label{eq:def-ZG}
\end{align}
Using the special point solution \eqref{eq:sol-sound}, $Z_G$ asymptotically behaves as
\begin{align}
   Z^\star_G
  &= C_1 (c^\star + \calI_\infty)\,
  \left[\, 1 + \frac{3}{2\, r} + \frac{3}{2\, r^2}
  + \frac{1}{2\, r^3}\, \left( 1 + \frac{3\, \sqrt{3}}{4}\,
        \frac{ e^{ - \frac{\pi}{ 2 \sqrt{3} } } }{ c^\star + \calI_\infty }
    \right) + O\big( r^{-4} \big)
  \, \right]
  ~.
\label{eq:def-Z_G-star}
\end{align}
Then, the Green's function at the special point depends on $c^\star$, and it is not unique there.

\section{Regularity from curvature invariants}\label{sec:curv_inv}

So far, we have studied regularity at the horizon from the behavior of gauge-invariant variables. In this section, we use curvature invariants to show regularity further. 

We consider pure gravity, so $R_{MN} \propto g_{MN}$. Then, the  curvature invariants $R_{MN}R^{MN}$ and $R^2$ remain unchanged under perturbations. Thus, we use the Kretschmann scalar $R_{KLMN}\, R^{KLMN}$. 
We thus consider the perturbed Kretschmann scalar 
\begin{align}
\delta(R_{KLMN}\, R^{KLMN}) := R_{KLMN}\, R^{KLMN}-\bmR_{KLMN}\, \bmR^{KLMN}~,
%
\end{align}
where the boldface letters indicate background values. But this quantity is not appropriate. Under the gauge transformation $x^M \to x^M + \xi^M$, a scalar $S$ transforms as 
\begin{align}
S \to S -\xi^M\del_M S~.
%
\end{align}
As a result, the perturbed Kretschmann scalar itself is not gauge invariant. 

The gauge-invariant Kretschmann scalar can be constructed in the same manner as gauge-invariant variables (see \appen{gauge_inv_variables}) and is given by
\begin{align}
\delta(R_{KLMN}\, R^{KLMN})_{\text{GI}} := \delta(R_{KLMN}\, R^{KLMN}) + \eta^r\del_r (\bmR_{KLMN}\, \bmR^{KLMN})~.
%
\end{align}
See \eq{eta_a-all} for $\eta^r$ which is constructed by $h_{MN}$. The explicit form of the gauge-invariant Kretschmann scalar can be found in \appen{curv_inv_app}.

First, assume $i\nw+\nq^4\neq0$, and use the master variable $\mfh_{rr}$. Earlier we found $(\lambda_1, \lambda_2) = (0, -2+i\nw)$. 
Using \eq{GI-Kretschmann-hrr}, the perturbed Kretschmann scalar becomes
\begin{align}
\delta(R_{KLMN}\, R^{KLMN})_{\text{GI}} 
\simeq -54a_0(\nw+i)(\nw+2i)~,
\quad (\text{for } \lambda_1)
%
\end{align}
so the $\lambda_1$-mode is indeed regular. On the other hand, 
\begin{align}
\delta(R_{KLMN}\, R^{KLMN})_{\text{GI}} 
\simeq \frac{54 a_0(i\nw-\nq^4)}{\nw(\nw+i)} (r-1)^{i\nw}~,
\quad (\text{for } \lambda_2)
%
\end{align}
Thus, the $\lambda_2$-mode is singular in general when $\Im \nw>0$. The $\lambda_2$-mode corresponds to the outgoing mode. We are interested in the retarded Green's function, so we are interested in the incoming mode, but the outgoing mode is prohibited from regularity as well. 

Note that we consider perturbations of the form $h_{MN}(r)e^{-i\omega v+iqx}$. Because the background spacetime is not flat, the perturbations contribute to the the perturbed Kretschmann scalar at $O(h_{MN})$. Thus, the above results (as well as expressions in \appen{curv_inv_app}) should be multiplied by $e^{-i\omega v+iqx}$. The Kretschmann scalar in real space is obtained by taking the real part. 

When $i\nw+\nq^4=0$, $\mfh_{rr}$ fails, and one cannot use the above result. At the special point $(\nw_\star, \nq_\star)$, one can use the exact solution \eqref{eq:sol-sound} and \eq{GI-Kretschmann-mfh0-simple}. One obtains 
\begin{align}
\delta(R_{KLMN}\, R^{KLMN})_{\text{GI}} 
\simeq 324\, C_1~.
%
\end{align}
Thus, the Kretschmann scalar remains regular at the horizon. Again, one should multiply the result by $e^{-i\omega_\star v+iq_\star x}$. But in this case, there is no need to take the real part. $\omega_\star$ and $q_\star$ are pure imaginary, but this just comes from the fact that we write the perturbation in plane-wave form. Recall that it actually behaves as $e^{\lambda(v-x/v_B)}$. Thus, the perturbations grow in time and so does the Kretschmann scalar. The perturbation breaks down eventually at late time. But the point is that the perturbation does not produce the diverging Kretschmann scalar near the horizon.
Note that the integration constant $C_2$ does not appear. In fact, the Kretschmann scalar is independent of $C_2$. 

One can obtain the general solution when $i\nw+\nq^4=0$ (footnote~\ref{fnote:solution}). When $i\nw=-1$ but $\nq^2\neq-1$, one can show that the Kretschmann scalar diverges at the horizon. Therefore, 
\begin{center}
Two solutions are regular only at the special point when $\Im\nw>0$. 
\end{center}
This also excludes the existence of any other special points in the upper-half $\omega$-plane.

\paragraph{Remarks}
In this paper, we show regularity using curvature invariants. 
Actually, it is difficult to show regularity of perturbations because there are many kinds of spacetime singularities. There are two common singularities:
\begin{itemize}
\item s.p.\ (scalar polynomial) curvature singularity
\item p.p.\ (parallelly propagated) curvature singularity
\end{itemize}
For a s.p.\ curvature singularity, curvature invariants diverge. 
For a p.p.\ curvature singularity, all curvature invariants remain finite%
\footnote{In the literature, some authors do not impose this condition for a p.p.\ singularity. Then,  s.p.\ singularities are part of p.p.\ singularities. }, 
but a tidal force diverges. Such singularities appear, for example, 
\begin{itemize}
\item in the extreme limit of some black $p$-branes and 
\item in the Lifshitz geometry \cite{Kachru:2008yh}. 
\end{itemize}
We have shown that 2 solutions do not produce a s.p.\ singularity at the special point, but strictly speaking, we have not shown that they do not produce a p.p\ singularity. 

A p.p.\ singularity is the one where the Riemann tensor components diverge in a p.p.\ frame along at least one non-spacelike curve. Physically, a radially infalling observer experiences a large tidal force. For example, 
\begin{enumerate}
\item Compute the Riemann tensor $R_{abcd}$ in a convenient orthonormal frame, usually in the static frame. 
\item An infalling observer measures the curvature not in the static frame but in another orthonormal frame which is related to the static frame by a local radial boost. 
\end{enumerate}


This procedure is a little complicated, but for our purpose, one does not need to carry out computations explicitly. The argument goes as follows:
\begin{enumerate}
\item[(i)] Suppose that our criterion holds, {\it i.e.}, 2 solutions are smooth at the future horizon. Then, the Riemann tensor in the EF frame is regular there. 
\item[(ii)] The magnitude of the Riemann tensor in general becomes large in the boosted frame (for a diagonal metric) \cite{Horowitz:1997uc}. But we use the incoming EF coordinates which is regular at the future horizon. Thus, the boost from the EF frame to the observer frame does not have a divergence at the horizon. 
\item[(iii)] Then, the regularity of the Riemann tensor in the EF frame implies the regularity of the Riemann tensor in the observer frame. Thus, 2 solutions do not produce a p.p.\ singularity.
\end{enumerate}

Nevertheless, we must stress that proving no p.p.\ singularity is very difficult. If the Riemann tensor diverges only along one curve, the spacetime is p.p.\ singular. In order to show that there is no p.p.\ singularity, one needs to examine the Riemann tensor along all curves, which is impossible in practice. 
What we can argue is that the perturbed spacetime is unlikely to have a p.p.\ singularity:
\begin{itemize}
\item First, as discussed above, $\mfh_{MN}$ is smooth at the horizon. 
\item Second, it is reasonable to focus on the radially infalling geodesic among all curves. 
Our interest is whether the outgoing mode is regular or not. The outgoing wave gets an infinite boost from the incoming wave point of view, so the radially infalling geodesic is likely to give the most strict condition.
\end{itemize} 

\section{Discussion}

In this paper, we examine the pole-skipping phenomenon using variables where regularity can be clearly seen. The Kretschmann scalar is also computed to show regularity. It is straightforward to analyze the special point if one uses the full set of gauge-invariant variables. This allows one to analyze the other systems where master variables are hard to find. However, dealing with the full set of equations is in general complicated. One way is to formulate the problem as an eigenvalue problem \cite{Natsuume:2019xcy}. 

We observed that the regular singularity at $r=1$ becomes a regular point at the special point in the incoming EF coordinates. As a result, two regular solutions appear. One would use this criterion to explore special points in the other systems \cite{Natsuume:2019xcy}. 

We show regularity using curvature invariants, but there are cases where one cannot use curvature invariants to show regularity. A simple example is the vector and the tensor modes of gravitational perturbations. For those modes, the perturbed Kretschmann scalar vanishes. We consider linear perturbations, so the quantities with different transformation properties decouple. The Kretschmann scalar transforms as a scalar, but the these modes transform differently. Thus, for those modes, the outgoing mode does not produce a s.p.\ singularity. However, it is likely that the outgoing mode (for a generic $\Im\nw>0$) is not smooth at the horizon. Then, it should produce a p.p\ singularity. 
In any case, the vector and tensor modes do not have special points in the upper-half $\omega$-plane \cite{Natsuume:2019xcy}, so the outgoing mode is not an issue there.

\section*{Acknowledgments}


We would like to thank Pavel Kovtun and Kengo Maeda for useful discussions. 
This research was supported in part by a Grant-in-Aid for Scientific Research (17K05427) from the Ministry of Education, Culture, Sports, Science and Technology, Japan. 


\appendix 

\section{Gauge-invariant variables}\label{sec:gauge_inv_variables}

We consider the background spacetime
\be
ds^2 = \bmg_{ab}(y) dy^a dy^b +e^{2\bmPhi} \delta_{ij}dx^i dx^j~,
%
\ee
where $y^a=(v,r)$, $x^i=(x,y)$%
\footnote{
For the quantities defined in the $p$-dimensional subspacetime ({\it e.g.}, $h^{(1)}_{ai}$ below), the index $i$ is raised and lowered with $\delta_{ij}$. For simplicity, we consider the $p$-dimensional metric which is proportional to $\delta_{ij}$, but the extention to $\gamma_{ij}(x)$ is easy. Replace $\del_i$ with $\bmD_i$, the covariant derivative with respect to $\gamma_{ij}$. Some expressions must be symmetrized since $\bmD_i$ do not commute.
}, 
and $e^{2\bmPhi}=r^2$. The 2-dimensional metric $\bmg_{ab}$ is given by
\begin{subequations}
\begin{align}
  & \bmg_{ab}
  = \begin{pmatrix} -r^2 f~ & 1 \\ 1 & 0 \end{pmatrix}
  ~,
 & \bmg^{ab}
  = \begin{pmatrix} 0 & 1 \\ 1~ & r^2 f \end{pmatrix}
  ~.
\label{A-eq:bg_metric} 
\end{align}
\end{subequations}
%

\subsection{Maxwell field example}

Let us start from the Maxwell field $A_M$. We assume that perturbations take the plane-wave form $A_M \propto e^{-i\omega v+ iqx}$. 
As we see below, the perturbations are decomposed as
\begin{align}
\text{scalar: }& A_v~, A_x~, A_r~, \\
\text{vector: }& A_y~.
%
\end{align}
%
%
It is not difficult to find variables which are invariant under the gauge transformation $A_M \to A_M-\del_M\lambda$. For the scalar mode, 
\begin{align}
\mfA_v &= A_v + \frac{\omega}{q}A_x~, 
\label{eq:inv_At} \\
\mfA_r &= A_r - \frac{1}{iq}A_x'~.
\label{eq:inv_Ar}
\end{align}
$\mfA_v$ and $\mfA_r$ are not independent: they are related by the Maxwell equation, and there is one master field for the scalar mode. 

It is not difficult to figure out the gauge-invariant variables for $A_M$: they are just proportional to the field strength $F_{MN}$. But for a systematic analysis, proceed as follows. 
We decompose the perturbations under the transformation of the boundary spatial coordinate $x^i$. The scalar (vector) mode transforms as scalar (vector) under the transformation. 
The Maxwell field consists of $A_M=(A_a, A_i)$. $A_i$ can be decomposed as 
\begin{align}
A_i = \del_iA_L+A_{T\,i}~, \quad
\del^i A_{T\,i} =0~.
\label{eq:dec-Ai}
\end{align}
The scalar mode consists of $A_a (A_v, A_r)$ and $A_L \propto A_x$, and the vector mode is $A_{T\,y}=A_y$. 

For $\lambda\propto e^{-i\omega v+ iqx}$, the gauge transformation 
\be
\delta A_M = -\del_M \lambda
%
\ee
becomes
\begin{subequations}
\begin{align}
\delta A_a &= -\del_a \lambda~, 
\label{eq:del_Aa} \\
\delta A_x &= iq \delta A_L = - iq \lambda~,\\
\delta A_{T\,i} &= 0~.
\label{eq:del_Ax}
\end{align}
\end{subequations}
Gauge-invariant variables eliminate the gauge parameter $\lambda$ by combining variables. The variables $A_{T\,i}$ are gauge invariant by themselves. From \eq{del_Ax}, the gauge parameter $\lambda$ is expressed by the perturbation $A_L$ as $\lambda = -\delta A_L$. Substituting this into \eq{del_Aa} gives
\be
\delta(A_a - \del_a A_L) = 0~,
%
\ee
so the gauge-invariant scalar perturbations are given by
\begin{align}
\mfA_a := A_a - \del_a A_L~.
\label{eq:inv_Aa}
\end{align}
In terms of components, \eq{inv_Aa} reduces to Eqs.~\eqref{eq:inv_At} and \eqref{eq:inv_Ar}. 

\subsection{Gauge-invariant metric perturbations}

Now consider metric perturbations $h_{MN} = (h_{ab}, h_{ai}, h_{ij})$. 
Again, the perturbations are decomposed as scalar, vector, and tensor mode. 
$h_{ab}$ gives 3 scalar perturbations. Just as the Maxwell field example, $h_{ai}$ is decomposed as
\begin{align}
h_{ai} = \del_i h_a + h^{(1)}_{ai}~, \quad
\del^i h^{(1)}_{ai} =0~,
\label{eq:dec-hai}
\end{align}
and $h_a$ gives 2 scalar perturbations and $h^{(1)}_{ai}$ gives vector perturbations. (The superscript ``$(1)$" refers to the spin.) In a similar manner, $h_{ij}$ is decomposed as
\begin{align}
h_{ij} =: h_L\, \delta_{ij} + \calP_{ij}\,h_T
  + 2 \del_{(i} h^{(1)}_{T\, j)} + h^{(2)}_{T\, ij}~,
\label{eq:dec-hij}
\end{align}
where
\begin{align}
  & 
  \del^i h^{(1)}_{T\, i} = 0~,
& & \del^j h^{(2)}_{T\, ij} = 0~,
& & h^{(2)}_{T\, i}{}^i = 0~,
%
\end{align}
and $\calP_{ij}$ is the projection operator given by
\begin{align}
\calP_{ij} := \del_i \del_j - \frac{1}{p} \delta_{ij}\, \del_k^2~.
%
\end{align}
The first term of $h_{ij}$ is the trace part which is a scalar perturbation. The rest is the traceless part which is decomposed as scalar $h_T$, vector $h^{(1)}_{T\, j}$, and tensor perturbations $h^{(2)}_{T\, ij}$. (For $p=2$, there is no tensor mode.) Thus, 
\begin{itemize}
\item The scalar mode consists of 7 perturbations $(h_{ab}, h_a, h_L, h_T)$. 
\item The vector mode consists of perturbations $(h^{(1)}_{ai}, h^{(1)}_{T\, i})$. 
\end{itemize}
In components (for $p=2$), the scalar mode is
\begin{align}
h_{xx} = h_L - \frac{1}{2}q^2 h_T ~, \quad
h_{yy} = h_L + \frac{1}{2}q^2 h_T ~, \quad
h_{vx} = iq h_v~, \quad
h_{rx} = iq h_r~,
%
\end{align}
and the vector mode is
\begin{align}
h_{ay} = h^{(1)}_{ay}~, \quad 
h_{xy} = iq h^{(1)}_{T\, y}~.
%
\end{align}

Again consider the gauge transformation $\delta_G x^M =\xi^M$. ($\delta_G$ refers to a gauge transformation.) The infinitesimal transformation $\xi^i$ is decomposed as
\begin{align}
\xi_i =: \del_i\xi_L+\xi_{T\,i}~, \quad
\del^i \xi_{T\,i} =0~.
\label{eq:dec-xi}
\end{align}
Only $\xi_a$ and $\xi_L$ appear for the scalar mode. 
The scalar mode transforms as
\begin{subequations}
\label{A-eq:del_G-h_M-all}
\begin{align}
  & \delta_G h_{ab} = - 2\, \twonabla_{(a} \xi_{b)}~,
\label{eq:del_hab} \\
  & \delta_G h_{a}
  = - \Big[~\xi_a + \partial_a \xi_L
  - 2\, \xi_L \big( \partial_a \bmPhi \big)~\Big]~,
\label{eq:del_ha} \\
  & \delta_G h_L
  = - \Big[~\xi^a\, \twonabla_a e^{2 \bmPhi}
  + \frac{2}{p}\, \del_k^2\, \xi_L~\Big]~,
\label{eq:del_hL} \\
  & \delta_G h_T = - 2\, \xi_L~,
\label{eq:del_hT} 
\end{align}
%
and the vector mode transforms as
%
\begin{align}
  & \delta_G h^{(1)}_{ai}
  = - \Big[~\partial_a \xi_{T\, i}
       - 2\, \xi_{T\, i} \big( \partial_a \bmPhi \big)~\Big]~,
\label{eq:del_hai} \\
  & \delta_G h^{(1)}_{T\, i} = - \xi_{T\, i}~,
\label{eq:del_hTi} 
\end{align}
\end{subequations}
where $\twonabla_a$ is the covariant derivative with respect to $\bmg_{ab}$. 

In order to obtain gauge-invariant variables, we again express gauge parameters $\xi_a$, $\xi_L$, and $\xi_{T\, i}$ by perturbations. For the vector mode, \eq{del_hTi} expresses $\xi_{T\, i}$ by $\delta_G h^{(1)}_{T\, i}$. Substituting \eq{del_hTi} into \eq{del_hai}, we obtain gauge-invariant vector perturbations:
\begin{align}
\delta_G \left( h^{(1)}_{ai} - \partial_a h^{(1)}_{T\, i}
  + 2\, h^{(1)}_{T\, i}~\partial_a \bmPhi  \right) = 0
\quad\to\quad
  \mfh_{ai}
  &:= h^{(1)}_{ai} - e^{2 \bmPhi}\,
    \partial_a \Big( e^{- 2 \bmPhi}\, h^{(1)}_{T\, i} \Big)~.
%
\end{align}
For the scalar mode, \eq{del_hT} expresses $\xi_L$ by $\delta_G h_T$. Substituting \eq{del_hT} into \eq{del_ha}, $\xi_a$ is expressed by $\delta_G h_a$ and $\delta_G h_T$:
\begin{subequations}
\label{A-eq:xi_a-all}
\begin{align}
  & \xi_a = \delta_G \eta_a
  ~,
\label{A-eq:xi_a} \\
  & \eta_a
  := \frac{1}{2}\, \partial_a h_T - h_T \partial_a \bmPhi - h_{a}~,
\label{eq:eta_a}
\end{align}
\end{subequations}
Substituting $\xi_a$ into \eq{del_hab}, we obtain
\begin{align}
\delta_G \left(h_{ab} + 2\, \twonabla_{(a} \eta_{b)} \right) = 0
\quad\to\quad
   \mfh_{ab}
  &:= h_{ab} + 2\, \twonabla_{(a} \eta_{b)}~.
\label{eq:inv_hab} 
\end{align}
Similarly, \eq{del_hL} becomes
\begin{align}
\delta_G \left( h_L + \eta^a\, \twonabla_a e^{2 \bmPhi}
  - \frac{1}{p}\, \del_i^2\, h_T \right) = 0
\quad\to\quad
   \mfh_L
  &:= h_L + \eta^a\, \twonabla_a e^{2 \bmPhi}
  - \frac{1}{p}\, \del_i^2\, h_T~.
\label{eq:inv_hL} 
\end{align}

Let us write these formulae in components. For $p=2$, the gauge-invariant vector perturbations are 
\begin{align}
\mfh_{vy} &= h_{vy} +\frac{\omega}{q} h_{xy}~, \\
\mfh_{ry} &= h_{ry} -\frac{r^2}{iq}\left( \frac{ h_{xy} }{r^2} \right)'~.
%
\end{align}
The gauge-invariant scalar perturbations are 
\begin{subequations}
\label{eq:scalar-mfh-all}
\begin{align}
   \mfh_{vv}
  &
  := h_{vv} - 2\, i\, \omega\, \eta_v + \bmg_{vv}' 
    \left( -\bmg_{vv}\, \eta_r + \eta_v \right)
  ~,
\label{A-eq:mfh_vv-I} \\
   \mfh_{vr}
  &:= h_{vr} + \eta_v'
  - \left( i\, \omega + \bmg_{vv}' \right)\, \eta_r
  ~,
\label{A-eq:mfh_vr-I} \\
   \mfh_{rr}
  &:= h_{rr} + 2\, \eta_r'
  ~,
\label{A-eq:mfh_rr} \\
   \mfh_L
  &
  = h_{yy}
  + 2r ( \eta_v - \bmg_{vv} \eta_r)~.
\label{A-eq:mfh_L-I}
\end{align}
\end{subequations}
From \eq{eta_a}, $\eta_a$ becomes 
\begin{subequations}
\label{eq:eta_a-all}
\begin{align}
   \eta_v
  &
  = - \frac{1}{i\, q} \left( h_{vx}
    + \frac{\omega}{q}\, \frac{ h_{xx} - h_{yy} }{2}
  \right)
  ~,
\label{A-eq:eta_v} \\
   \eta_r
  &
  = \frac{1}{i\, q}\, \left\{ \frac{ r^2 }{i\, q}\,
       \left( \frac{ h_{xx} - h_{yy} }{2\, r^2} \right)'
    - h_{rx} \right\}
  ~.
\label{A-eq:eta_r}
\end{align}
\end{subequations}
%

\section{Some formulae}

\subsection{Expressions of gauge-invariant variables by $\mfh_{rr}$}\label{sec:relation_mfh}

%
\begin{subequations}
\label{eq:mfh-by-hrr-all}
\begin{align}
  \frac{ \mfh_{vv} }{3}
  & = \frac{1}{4} \frac{r^5}{ i \mfw + \mfQ^4 }~
  \left( P_{vv}(r) f r \odiff{}{r}
    + 3 Q_{vv}(r) \right) \mfh_{rr}
  ~,
\label{A-eq:mfh_vv-by-h_rr} \\
  & P_{vv}
  := i \mfw \left\{ \mfw^2 + \frac{1}{r}
  - \frac{f}{3} \left( 2 \mfQ^2 + \frac{1}{r} \right) \right\}
  ~,
\label{A-eq:def-P_vv} \\ 
  & Q_{vv}
  := i \mfw \left( \mfw^2 + \frac{1}{r} \right)
    \left( 2 - \frac{i \mfw}{r} \right)
  + i \mfw \frac{f}{3} \left\{
     3 i \mfw \left( i \mfw - 1 + \frac{f}{3} \right)
     - \frac{4 + 3 \mfw^2}{r} + \frac{2 i \mfw}{r^2}
  \right\}
\nonumber \\
  &\hspace*{1.0truecm}
  + 3 i \mfw (i \mfw - \mfQ^2) \frac{f}{3}
    \left( 1 - \frac{i \mfw}{r} - \frac{f}{3} \right)
  + 2 \frac{ i \mfw + \mfQ^4 }{r} \frac{f^2}{9}
  ~.
\label{A-eq:def-Q_vv} \\
%
%
  \mfh_{vr}
  &= - \frac{r^2 f}{2} \mfh_{rr}
  ~.
\label{A-eq:mfh_vr-by-h_rr} \\
%
%
  \frac{ \mfh_{L} }{ - 2 r }
  &= \frac{1}{4} \frac{r^5}{ i \mfw + \mfQ^4 }~
  \left( P_{L}(r) f r \odiff{}{r}
    + 3 Q_{L}(r) \right) \mfh_{rr}
  ~,
\label{A-eq:mfh_L-by-h_rr} \\
  & P_L
  := (i \mfw - \mfQ^2) \left( 1 - \frac{i \mfw}{r} \right)
  - i \mfw \left( 1 - \frac{1}{r} \right)
    \left( 1 + i \mfw + \frac{1}{r} \right)
  + \mfQ^2 \frac{f}{3}
  ~,
\label{A-eq:def-P_L} \\
  & Q_L
  := (i \mfw - \mfQ^2)
    \left( 2 - \frac{i \mfw}{r} \right)
    \left( 1 - \frac{i \mfw}{r} \right)
  - i \mfw \left( 1 - \frac{1}{r} \right)
    \left( 2 + 3 i \mfw + \frac{2 + \mfw^2}{r} \right)
\label{A-eq:def-Q_L} \\
  &\hspace*{1.0truecm}
  + \frac{f}{3} \left\{
    i \mfw \left( 2 - \frac{1}{r} \right)
    \left( 2 + 3 i \mfw + \frac{1}{r} \right)
  - (i \mfw - \mfQ^2) \left( 4 - 3 \frac{i \mfw}{r} \right)
  - \frac{ i \mfw + \mfQ^4 }{r^2} \right\}
  ~.
\nonumber
\end{align}
\end{subequations}
Note that $\mfh_{vv}$ and $\mfh_L$ are proportional to $(i\nw+\nq^4)^{-1}$, and it diverges at $i\nw+\nq^4=0$ which includes the special point $(\nw,\nq)=(i,\pm i)$. Thus, the master variable $\mfh_{rr}$ fails at the special point, and the special point must be examined separately. 

\subsection{Field equations near the special point}
\label{sec:perturbed_eqs_app}

%
\begin{subequations}
\label{eq:exp-EOM-mfh-sp}
\begin{align}
   0
  &= 2 r^2 \delta\mfh_{vv}
  - (r^2 + r + 1)^3 \delta\mfh_{rr}
  - 3 (r + 1) \delta\mfh_L
\nonumber \\
  &\hspace*{0.5truecm}
  + \frac{ 27  i \delta\mfw\, r (r + 1) }
         { (r - 1)^3 (r^2 + r + 1) }
    \Bigg[ - \frac{ 2 r^2 (3 r^3 + r^2 + r + 1) }
                    { 9 (r + 1) }
       \left\{ \frac{\delta\eta}{i \delta\mfw}
        + \frac{ r (r - 1) (2 r + 1) }
               { 3 r^3 + r^2 + r + 1 }
      \right\} \mfh^\star_{vv}
\nonumber \\
  &\hspace*{1.0truecm}
  + \left\{ 1 + \frac{ r^3 f }{9 (r + 1) }
      \left( 7 + 7 r + r^2
        + \frac{\delta\eta}{i \delta\mfw}
          \frac{ (2 r + 1) (2 r^3 + 3 r^2 + 1) }
               { r (r^2 + r + 1) } \right)
    \right\} \mfh^\star_L
   \Bigg]
  ~,
\label{eq:exp-EOM-const1-sp} \\
   0
  &= \delta\mfh'_{vv}
  - \left( 1 - \frac{1}{r } \right)
    \frac{ \delta\mfh_{vv} }{r}
  + \frac{3}{2} \left( \frac{1}{2} - \frac{f}{3} \right)
    r^2 f \delta\mfh_{rr}
  - \frac{3}{2 r^4} \delta\mfh_L
\nonumber \\
  &\hspace*{0.5truecm}
  + i \delta\mfw \left\{
    \frac{\delta\eta}{i \delta\mfw} \frac{ \mfh^\star_{vv} }{r^2}
  - \frac{3}{2} \left( \frac{1}{2}
    + \frac{\delta\eta}{i \delta\mfw} \frac{f}{3} \right)
    r^2 f \mfh^\star_{rr}
  \right\}
  ~,
\label{eq:exp-EOM-hvv-sp} \\
   0
  &= \delta\mfh'_{rr}
  + 3 \frac{ 1 + r }{ 1 + r + r^2 } \delta\mfh_{rr}
\nonumber \\
  &\hspace*{0.5truecm}
  + \frac{ 27 i \delta\mfw (r^2 + 2 r + 3) }
         { r^{10} f^4 }
    \Bigg[ - \frac{ 2 (r^2 + 1) (r^2 + r + 1) }
                    { 3 (r^2 + 2 r + 3) }
      \left\{ \frac{\delta\eta}{i \delta\mfw}
        + \frac{ r^3 (3 r^3 + r^2 + 4 r + 1) }
               { 3 (r^2 + 1) (r^2 + r + 1)^2 } f
      \right\} \mfh^\star_{vv}
\nonumber \\
  &\hspace*{1.0truecm}
    + \left\{ 1 + \frac{ 11 r^3 - 3 r - 2 }
                       { 9 (r^2 + 2 r + 3) }
    \left( 1 + \frac{\delta\eta}{i \delta\mfw}
          \frac{ 5 r^3 + 6 r^2 + 3 r + 4 }
               { 11 r^3 - 3 r - 2 } \right) f
      \right\} \mfh^\star_{L}
   \Bigg]
  ~,
\label{eq:exp-EOM-hrr-sp} 
\displaybreak \\
   0
  &= \delta\mfh'_{L}
  - \frac{1}{2} \left( 1 + \frac{3 r^2}{r^2 + r + 1} \right)
     \frac{ \delta\mfh_{L} }{r}
  - \frac{r}{r^2 + r + 1} \delta\mfh_{vv}
  - \frac{1}{2} r^2 f \delta\mfh_{rr}
\nonumber \\
  &\hspace*{0.5truecm}
  + \frac{3 i \delta\mfw}{2 r^2 f} \Bigg[
      \frac{\delta\eta}{i \delta\mfw}
      \frac{2 r}{r^2 + r + 1} \mfh^\star_{vv}
    + \left\{ -1 + \frac{\delta\eta}{i \delta\mfw}
      \frac{ r^4 (r + 1) f^2 }{ (r^2 + r + 1)^3 } \right\}
       \mfh^\star_{L}
   \Bigg]
  ~.
\label{eq:exp-EOM-hL-sp}
\end{align}
\end{subequations}
As discussed in the main text, we require that the source terms are written as Taylor series. Note that the source term of \eq{exp-EOM-hvv-sp} does not contain $(r-1)^{-1}$, so the equation gives no condition. 

\subsection{Gauge-invariant perturbed Kretschmann scalar}\label{sec:curv_inv_app}

The gauge-invariant Kretschmann scalar is given by
%
\begin{subequations}
\label{A-eq:GI-Kretschmann_scalar-mfh-all}
\begin{align}
  & \delta_{\text{GI}} (R_{KLMN} R^{KLMN})
\nonumber \\
  &= \frac{12}{ i \mfw r^7 } \left\{
    2 \mfQ^2 r \mfh_{vv}
  + 3 i \mfw \left( \mfQ^2 - \frac{1}{r} \right)  \mfh_{L}
  - \mfQ^2 r^5 f^2 \mfh_{rr}
  \right\}
\label{eq:GI-Kretschmann-mfh0-simple} \\
  &= \frac{- 18}{r^2}
  \bigg[ \left( i \mfw - \frac{1}{r^2} \right)
    r f \mfh_{rr}'
  + 3 \left\{ \left( i \mfw - \frac{2}{r^2} \right)
    \left( 1 - \frac{i \mfw}{r} \right)
    - \frac{f}{3 r} \left( \mfQ^2 - \frac{2}{r} \right)
      \right\} \mfh_{rr}
   \bigg]
  ~.
\label{eq:GI-Kretschmann-hrr}
\end{align}
\end{subequations}
In the first expression, we write the scalar only by $\mfh_{MN}$ eliminating $\mfh'_{MN}$ and $\mfh''_{MN}$ using field equations. The expression is not unique because of the constraint equation \eqref{eq:EOM_const1}. In the last expression, we write it by $\mfh_{rr}$ and $\mfh'_{rr}$ again using field equations. However, it uses expressions such as \eq{mfh-by-hrr-all}, but they are valid only when $i\nw+\nq^4 \neq 0$.


\end{document}